# Strategic Teaching and Learning in Games

Burkhard C. Schipper*

April 12, 2015

**Abstract**

It is known that there are uncoupled learning heuristics leading to Nash equilibrium in all finite games. Why should players use such learning heuristics and where could they come from? We show that there is no uncoupled learning heuristic leading to Nash equilibrium in all finite games that a player has an incentive to adopt, that would be evolutionary stable or that "could learn itself". Rather, a player has an incentive to strategically teach such a learning opponent in order secure at least the Stackelberg leader payoff. The impossibility result remains intact when restricted to the classes of generic games, two-player games, potential games, games with strategic complements or $2 \times 2$ games, in which learning is known to be "nice". More generally, it also applies to uncoupled learning heuristics leading to correlated equilibria, rationalizable outcomes, iterated admissible outcomes, or minimal curb sets. A possibility result restricted to "strategically trivial" games fails if some generic games outside this class are considered as well.

**Keywords:** Learning in games, learning heuristics, learning rules, interactive learning, uncoupled learning, meta-learning, Nash equilibrium, correlated equilibrium, rationalizability, iterated admissibility, minimal curb sets, dominance solvable games, common interest games, reputation.

**JEL-Classifications:** C72, C73.

*Department of Economics, University of California, Davis, Email: bcschipper@ucdavis.edu; Part of the research was undertaken when the author visited New York University and the University of Heidelberg. I thank Yakov Babicheko, Jerney Copic, Peter Duersch, Drew Fudenberg, Hans Haller, Martin Meier, Ichiro Obara, Joerg Oechssler, Joe Ostroy, Tomasz Sadzik, Klaus Schmidt, and Dale Stahl as well as seminar audiences at Queen's University, UC Davis, UCLA, USC, UT Austin, U Wisconsin-Madison, Virginia Tech, and SAET Paris 2013 for helpful discussions. Financial support through NSF CCF-1101226 is gratefully acknowledged.

# 1 Introduction

Individuals are not born with complete knowledge but with an ability to learn. In economic and social situations, learning is interactive as the behavior of learners is a part of what learning is about. There is now a large literature that studies how individuals learn interactively equilibrium (see for an overview Fudenberg and Levine, 1998, or Young, 2004). More recently, this literature focused on *uncoupled* learning heuristics that if followed by all players lead to a stage-game equilibrium in *all* games (e.g., Foster and Young, 2003, 2006, Hart and Mas-Colell, 2003, 2006, Germano and Lugosi, 2007, Kakade and Foster, 2008, Young, 2009). A learning heuristic is uncoupled if it does not depend on the opponents' payoff functions. An argument for *uncoupledness* is that opponents' preferences are usually not directly observable. Thus, learning should "work" also without knowledge of the opponents' payoffs. The desire to find learning heuristics that lead to equilibrium in *all* games is perhaps motivated by the fact that learning is present in all social or economic situations. Consequently, it is tempting to find general features of learning that makes it "work" in all situations rather than just learning heuristics tailored in an ad hoc fashion to particular situations. Given the complications of learning in interactive contexts, it may look hopeless to find a learning heuristic that both is uncoupled and reaches equilibrium in all games. Indeed, there are various impossibility results (Nachbar, 1997, 2001, 2005, Foster and Young, 2001, Hart and Mas-Colell, 2003, Sadzik, 2011). Yet, there is also a stream of positive results. Some of the uncoupled learning heuristics that lead to stage-game Nash equilibrium in all finite games are surprisingly simple (e.g., Foster and Young, 2003, Foster and Young, 2006, Hart and Mas-Colell, 2006, Germano and Lugosi, 2007, Kakade and Foster, 2008, Young, 2009) while other learning heuristics involve sophisticated information processing, optimization, and suitable initial beliefs (e.g., Jordan, 1991, Kalai and Lehrer, 1993). These positive results are paramount for justifying the wide-spread use of equilibrium in economic analysis. The application of equilibrium to one-shot strategic situations can be motivated by the agents' ability to learn equilibrium when these situations would be repeated.

In this paper, we raise the following question: Suppose that players adopt uncoupled learning heuristics that lead to stage-game equilibrium in all games. Why should a player have an individual incentive to follow such a learning heuristic? More precisely, are there uncoupled learning heuristics that lead to stage-game equilibrium in all finite games *and* that every player has an incentive to adopt if all other players adopt it as well? The answer to this question is relevant for at least three reasons: First, the simpler learning heuristics leading to Nash equilibrium, the more confident we are that people are able to reach equilibrium. Yet, it is hard to imagine that all people in reality are oblivious to the fact that they strategically interact. They may also be strategic when it comes to learning. A player may learn to some extent about the opponents' learning and use this knowledge to deviate from her current learning heuristic in order to *strategically teach* the opponents. Strategic teaching would not call into question the learning foundation for equilibrium if it always enhances learning or selection of equilibrium (as in instances of the experimental study by Hyndman, Ozbay, Schotter, and Ehrblatt, 2012, or in the theoretical study with a slightly different set-up by Ellison, 1997). Yet, strategic teaching may conflict with strategic learning of equilibrium if a player uses strategic teaching of opponents to her individual advantage leading them play further away from stage-game equilibrium (as suggested in experimental studies by Camerer, Ho, and Chong, 2002, Duersch, Kolb, Oechssler, and Schipper, 2010, and Terracol and Vaksmann, 2009). A sufficient condition for learning being robust to individually beneficiary strategic teaching attempts is that players want to adopt the learning heuristic if opponents adopt it as well.

The second reason for studying incentives of adoption of learning heuristics is evolution. If basic abilities to learn are biologically or culturally heritable[1], then it is meaningful to consider general learning heuristics that in some sense are *evolutionary stable*. That is, consider a game in which players are programmed to learning heuristics that they use to learn how to play in games and payoffs are the long run payoffs from profiles of learning heuristics "averaged" over all games. Call this the learning game. A necessary condition for a learning heuristic to satisfy reasonable notions of evolutionary stability is that it is a best response to itself in the learning game.

The final reason for studying incentives for the adoption of learning heuristics is learning to learn. Players may not just learn to play games but also learn about how to learn to play games. Learners may reflect upon their learning and adjust how they learn. This has been called *meta-learning* in social, cognitive, and educational psychology (e.g., Biggs, 1985). Similarly, the term meta-learning has also been used in computer science and artificial intelligence to refer to learning algorithms that improve themselves. Finding such algorithms has been called the "grand problem of artificial intelligence" (Schmidhuber, 2003). Learning to learn may not stop at the second level. In principle, we could consider also players who learn about how to learn about how to learn to play games, learn about that ... ad infinitum. The question emerges whether there is a *universal learning heuristic* to would solve the infinite regress problem of learning. Since we require the uncoupled learning heuristic to lead to Nash equilibrium in all games, a necessary condition for a learning heuristic to be able to learn itself would be that the learning heuristic is a Nash equilibrium of the learning game. We show that the answers to all questions above are negative. That is, there is no profile of uncoupled learning heuristics that leads to Nash equilibrium in all finite games and that players have an incentive to adopt if opponents follow it. More generally, this impossibility result applies not only to uncoupled learning heuristics leading to Nash equilibrium but also uncoupled learning heuristics approaching the set of correlated equilibria, the set of rationalizable action profiles, the set of iteratively admissible action profiles, or minimal curb sets. Moreover, the impossibility result applies even if instead of all games we just consider "nice" subclasses of games for which many learning dynamics are known to converge to equilibrium such as $2 \times 2$ games, ordinal potential games, or games with strategic complementarities etc. Showing the impossibility is surprisingly simple by constructing an appropriate counterexample.

---

[1]Recently, Rietveld et al. (2013) reported on a robust but small genetic association with educational attainment. Moreover, Caprara et al (2008) reported on a study with high-school students showing that indeed academic achievement depends on learning strategies employed. Taken together, this evidence suggests to us that biological heritability of learning abilities may not be a far-fetched idea.

What can a player gain from strategic teaching when he faces an opponent who follows an uncoupled learning heuristic leading to Nash equilibrium in all games? We show that she can gain at least the "average" over all games of the Stackelberg leader payoffs and that this is strictly more than adopting an uncoupled learning heuristic that would allow both players to reach Nash equilibrium in all games. This is reminiscent of the reputation results in repeated games (see Mailath and Samuelson, 2006, Part IV, for a review). For instance, Fudenberg and Levine (1989) consider repeated games with a long-run player who faces a sequence of short-run players. Short-run players are uncertain about the payoff-type of the long-run player and thus the game they are playing. In particular, there is strict positive prior probability that the long-run player is of the payoff-type for which the Stackelberg leader strategy of the "true" game is strict dominant. They show that the long-run player's payoff converges the Stackelberg leader payoff as she becomes sufficiently patient. In our setting, there is no Bayesian game. Yet, the role of uncertainty is taken by uncoupledness of the learning heuristic as it means that the learning heuristic cannot distinguish between "payoff-types" of the other player. Implicitly there are many "payoff-types" of the other player in our setting since we require an uncoupled learning heuristic to face the other player in *all* games (because it is supposed to converge to Nash equilibrium in all games or at least in special but rich subclasses of games). The long-run player corresponds now to the strategic teacher who cares about her long-run payoff that she can achieve against an opponent following an uncoupled learning heuristic leading to Nash equilibrium in all games. Moreover, the opponent following an uncoupled learning heuristic leading to Nash equilibrium in all games is short-term since he implicitly cares about converging to *stage-game* Nash equilibrium in all games. To some extent, Fudenberg and Levine (1998, Chapter 8.11.1) anticipated our observations in the last chapter of their book on "The Theory of Learning in Games" when they suggested that reputation results from the repeated games literature should carry over to the learning literature. Yet, they had a very different setting in mind in which they explicitly added uncertainty. Moreover, they did not show that any uncoupled learning heuristic that leads to Nash equilibrium provides an opportunity for strategic teaching and developing reputations.

What about positive results? Our counterexample begs the question whether there is a "maximal" class of games for which there are uncoupled learning heuristics that lead to Nash equilibrium in all games of this class, that each player has an incentive to adopt if opponents adopt their part, and for which this possibility fails the moment some other games outside the class are considered as well. We show that when we restrict to the class of games that can be solved by one round of elimination of weakly dominated actions, then we obtain such a possibility result. I.e., for the class of one-round dominance solvable games there exists an uncoupled learning heuristic that leads to Nash equilibrium in all games of this class and that each player has an incentive to adopt. But this class is also maximal in the sense that if we consider a generic game beyond this class as well, then we obtain a negative result again. Note that one-round dominance solvable games are in some sense "strategically trivial" since each player could play optimally when treating the game as a single-person decision problem. We show a similar possibility result for common interest games, i.e., games that possess a payoff profile that strictly Pareto dominates any other payoff profile. In such games, each player's maximal payoff identifies a pure Nash equilibrium.

The paper is structured as follows: The next section outlines the model. The counterexample is presented in Section 3. The generality of the example is explored in Section 4. In Section 5 we establish lower bounds on payoffs achievable with strategic teaching. Some "possibility" results are presented in Section 6. Finally, we conclude with a discussion in Section 7. All proofs are fortunately elementary and collected in the appendix.

# 2 Model

For our purpose it will turn out to be enough to consider two-player games with players Rowena $R$ and Colin $C$. For simplicity, each player has the same nonempty finite set of actions $A$. As customary, $i \in \{R, C\}$ refers to one player and $-i \in \{R, C\}$ refers to $i$'s opponent. The payoff function of player $i$ is denoted by $u_i : A \times A \longrightarrow [0, 1]$. We normalize payoffs to be in the unit interval for integrability reasons. Consider now the class of *all* normal-form two-player games with the action set $A$. The space of all such games is generated by varying players' payoff functions. Since $A$ is finite, it is identified by the $2|A^2|$ - dimensional Euclidian vector space $[0, 1]^{2|A^2|}$. Thus, we may simply call a profile of payoff functions $\mathbf{u} = (u_R, u_C) \in [0, 1]^{2|A^2|}$ a game. Let $\lambda$ be the Lebesgue measure on the space of games $[0, 1]^{2|A^2|}$. The Lebesgue measure gives us a notion of "size" of classes of games.

For each game $\mathbf{u} \in [0, 1]^{2|A^2|}$, the dynamic setup consists of a repeated play of the stage-game $\mathbf{u}$ at discrete time periods $t = 0, 1, 2, \ldots$. Let $a_i^t \in A$ denote the action of player $i$ at time $t$, and $\mathbf{a}^t = (a_R^t, a_C^t) \in A \times A$ be the combination of actions at $t$. At the end of period $t$ each player $i$ observes the combination of realized actions $\mathbf{a}^t$. That is, we assume perfect monitoring of play.

A learning heuristic $\sigma_i$ of player $i$ assigns to each game $\mathbf{u} \in [0, 1]^{2|A^2|}$ a sequence of functions $(\sigma_i^0(\mathbf{u}), \sigma_i^1(\mathbf{u}), \ldots, \sigma_i^t(\mathbf{u}), \ldots)$ where for each $t > 0$ the function $\sigma_i^t(\mathbf{u})$ assigns to each history $h^{t-1} := (\mathbf{a}^0, \mathbf{a}^1, \ldots, \mathbf{a}^{t-1})$ in the $t$-th repetition of the stage game $\mathbf{u}$ a mixed action[2] in $\Delta(A)$ (the set of probability distributions on $A$) to be played at stage $t$. With this formulation we let $\sigma_i^0(\mathbf{u})$ be simply player $i$'s distribution of initial actions in game $\mathbf{u}$ when $i$ follows learning heuristic $\sigma_i$. We assume that each $\sigma_i^t$, $t = 0, 1, \ldots$, is Lebesgue measurable with respect to the space of games. We denote by $\boldsymbol{\sigma} = (\sigma_R, \sigma_C)$ a profile of learning heuristics. The set of learning heuristics is denoted by $\Sigma$.[3]

---

[2] Allowing for mixed actions and thus stochastic learning heuristics is crucial (even for learning pure Nash equilibrium). Hart and Mas-Colell (2003) show that there exist no *deterministic* uncoupled learning heuristics converging to Nash equilibrium in all games while Hart and Mas-Colell (2006) show that there are *stochastic* uncoupled learning heuristics that converge to Nash equilibrium in all games. Randomization of actions allows for exhaustive search.

[3] Our notion of learning heuristics allows for any measurable behavioral strategy. This is in line with some of the learning literature (e.g., Hart and Mas-Colell, 2006). More restrictively, we could consider learning heuristics to be finite automatons as in Babichenko (2010). Yet, we believe that opting for the more general notion strengthens our negative result according to which there is no learning heuristic satisfying certain conditions. Alternatively, we could take learning heuristics as "procedures" that by the Church-Turing hypothesis may be formalized by Turing machines. Yet, the learning game won't be well-defined when a player uses a Turing machine that doesn't halt. Moreover, the cardinality of the set

Following Hart and Mas-Colell (2003, 2006), we consider uncoupled learning heuristics. These learning heuristics may take opponents' actions and the player's payoff function as an input but not opponents' payoff functions.

**Definition 1 (Uncoupled)** *A learning heuristic $\sigma_i$ is uncoupled if for all $(u_i, u_{-i}) \in [0,1]^{2 \cdot |A^2|}$, $\sigma_i(u_i, u_{-i}) = \sigma_i(u_i, \hat{u}_{-i})$ for all $\hat{u}_{-i} \in [0,1]^{|A^2|}$.*

The following notion of convergence can be viewed both as strong and weak. It is strong because we require almost sure convergence.[4] We believe that it captures best what we intuitively mean with "convergence". Yet, below version is also a weak notion of convergence in the sense that we require convergence to pure equilibrium only and are silent on convergence in games with mixed equilibrium only. As it will become clear later, the focus on pure equilibrium will be enough for our purpose.

**Definition 2 (Convergence to Nash Equilibrium)** *A profile of learning heuristics $\boldsymbol{\sigma}$ converges to Nash equilibrium in every stage-game if for every game $\mathbf{u} \in [0,1]^{2|A^2|}$ that possesses a pure Nash equilibrium, almost every play path consists of a pure stage-game Nash equilibrium being played from some point on. We say that learning heuristic $\sigma$ converges to Nash equilibrium in every stage-game if it is a component of a profile of learning heuristics that converges to Nash equilibrium in every stage-game.*

Next we describe a game in which the "actions" are the learning heuristics. Later in the text, we abuse slightly notation and let $u_i$ also denote the multilinear extended (expected) utility function defined on the space of mixed action profiles $\Delta(A) \times \Delta(A)$. We denote by $\mathbf{a}^t(\boldsymbol{\sigma}(\mathbf{u}))$ a profile of actions realized at period $t$ under a history generated by the profile of learning heuristics $\boldsymbol{\sigma}$ in the repeated stage-game $\mathbf{u}$. For each profile of learning heuristics $\boldsymbol{\sigma}$ and each game $\mathbf{u} \in [0,1]^{2|A^2|}$, we denote player $i$'s limit of expected means payoff by

$$v_i(\boldsymbol{\sigma}(\mathbf{u})) := \lim_{T \to \infty} \inf \mathbb{E}_{\boldsymbol{\sigma}(\mathbf{u})} \left[ \frac{1}{T} \sum_{t=1}^{T} u_i(\mathbf{a}^t(\boldsymbol{\sigma}(\mathbf{u}))) \right], \tag{1}$$

---

of Turing machines is smaller than the cardinality of the set of behavioral strategies. (See Megiddo and Wigderson (1986) and Megiddo (1986) for discussions of games with Turing machines.)

[4]Some of our observations generalize to some weaker notions of convergence. To keep the exposition simple and conceptually straightforward, we use almost sure convergence.

where the expectations are formed over mean payoffs[5] resulting from histories given positive probability by the profile of learning heuristics $\boldsymbol{\sigma}$ in the repeated stage-game $\mathbf{u}$. This is the long run expected payoff to player $i$ in game $\mathbf{u}$ emerging from the profile of learning heuristics $\boldsymbol{\sigma}$. Note that $v_i$ is a measurable random variable on $\mathbf{U}$. To see this note that by assumption $\sigma_i^t$ is measurable for every $t = 0, 1, \ldots$. Further, for every $T$, $\mathbb{E}_{\boldsymbol{\sigma}(\mathbf{u})}\left[\frac{1}{T}\sum_{t=1}^{T} u_i(\mathbf{a}^t(\boldsymbol{\sigma}(\mathbf{u})))\right]$ is linear in probabilities of the per-period behaviorial strategies on a finite dimensional real-valued domain. Hence it is continuous in those probabilities and thus measurable on $\mathbf{U}$. Since the lim inf of a sequence of measurable real-valued functions is measurable, we have that $v_i$ is a measurable random variable on $\mathbf{U}$.

We are interested in the long run expected payoff "averaged" over all games in Lebesgue measurable subsets of games $\mathbf{U} \subseteq [0,1]^{2|A^2|}$ defined by

$$V_i(\boldsymbol{\sigma}, \mathbf{U}) := \int_{\mathbf{U}} v_i(\boldsymbol{\sigma}(\mathbf{u})) d\lambda. \tag{2}$$

This defines a normal-form game $\langle \{R, C\}, \Sigma, (V_i(\cdot, \mathbf{U}))_{i=R,C} \rangle$ in which each player $i$ "chooses" a learning heuristic in $\Sigma$ and her payoff from a profile of learning heuristics $\boldsymbol{\sigma} \in \Sigma \times \Sigma$ over all games in $\mathbf{U}$ is given by $V_i(\boldsymbol{\sigma}, \mathbf{U})$. We call this game the *learning game* based on $\mathbf{U}$.

We are interested in learning heuristics that are Nash equilibrium of a learning game. Existence of Nash equilibrium of the learning game is guaranteed. There is a learning heuristic that for each stage-game $\mathbf{u} \in \mathbf{U}$ prescribes a Nash equilibrium of this stage-game. Of course, such a strategy would not necessarily be uncoupled except when considering just some special classes of games.

**Definition 3 (Nash Equilibrium of the Learning Game)** *A profile of learning heuristics $\boldsymbol{\sigma} = (\sigma_R, \sigma_C) \in \Sigma \times \Sigma$ is a Nash equilibrium of the learning game $\langle \{R, C\}, \Sigma, (V_i(\cdot, \mathbf{U}))_{i=R,C} \rangle$ if*

$$V_i(\sigma_i, \sigma_{-i}, \mathbf{U}) \geq V_i(\hat{\sigma}_i, \sigma_{-i}, \mathbf{U}) \text{ for all } \hat{\sigma}_i \in \Sigma.$$

As we explained in the introduction, we view that requiring the learning heuristic to be a Nash equilibrium action of the learning game is a weak necessary condition for a strategic, evolutionary, or learning foundation of learning heuristics.

---

[5] We choose the limit of expected means payoff over alternative ways to evaluate streams of payoffs because we are interested what payoffs learning heuristics achieve in the limit when they had a fair chance to learn equilibrium. We conjecture that analogous results obtain with discounting when players are sufficiently patient.

# 3  A Simple Counterexample

Consider the following $2 \times 2$ game $\mathbf{u}^1$

$$\begin{array}{c} & \begin{array}{cc} a & b \end{array} \\ \begin{array}{c} b \\ c \end{array} & \left( \begin{array}{cc} 16,12 & 13,13 \\ 17,7 & 14,6 \end{array} \right) \end{array}$$

In this game, action $c$ of Rowena strictly dominates her action $b$. The unique Nash equilibrium of the game is $(c, a)$, which is in pure actions. If players follow uncoupled learning heuristics converging to Nash equilibrium in all finite games, then they must reach $(c, a)$ in game $\mathbf{u}^1$.

Next consider the $2 \times 2$ game $\mathbf{u}^2$

$$\begin{array}{c} & \begin{array}{cc} a & b \end{array} \\ \begin{array}{c} b \\ c \end{array} & \left( \begin{array}{cc} 16,12 & 13,13 \\ 15,7 & 9,6 \end{array} \right) \end{array}$$

This game is identical to game $\mathbf{u}^1$ except for the payoffs of Rowena from action $c$. Now $b$ strictly dominates action $c$. The unique Nash equilibrium is $(b, b)$, which again is in pure actions. Any profile of learning heuristics that is uncoupled and converges to Nash equilibrium in all games must convergence to $(b, b)$.

Would Rowena have an incentive to stick to such a learning heuristic? Note that if Rowena behaves in game $\mathbf{u}^2$ exactly as in game $\mathbf{u}^1$, then since the Colin follows an uncoupled learning heuristics the play must converge to $(c, a)$ in game $\mathbf{u}^2$. This would yield Rowena a payoff of 15, which is strictly larger than her payoffs of 13 in the unique Nash equilibrium of $\mathbf{u}^2$. Thus, Rowena has a strict incentive to deviate from an uncoupled learning heuristic converging to Nash equilibrium in all games to a "strategic teaching" heuristic as just described.

Since both games are generic, we can consider open neighborhoods $\mathbf{U}^1$ and $\mathbf{U}^2$ of $\mathbf{u}^1$ and $\mathbf{u}^2$, respectively, such that both $\mathbf{U}^1$ and $\mathbf{U}^2$ belong to the class of games $\mathbf{U}$ for which there exists a pure Nash equilibrium. Let $(\sigma_R, \sigma_C)$ be a profile of uncoupled learning heuristic that converges to Nash equilibrium in games in $\mathbf{U}$. Moreover, let $\sigma'_R$ be a learning heuristic that plays both in games in $\mathbf{U}^1$ and $\mathbf{U}^2$ identical to $\sigma_R$ in $\mathbf{U}^2$ and identical to $\sigma_R$ in all other games in $\mathbf{U}$. Since the space of games $\mathbf{U}$ contains the nonempty open sets $\mathbf{U}^1$ and $\mathbf{U}^2$ and the Lebesgue measure is strictly positive on nonempty open subsets of the space of games[6], the learning heuristic $\sigma'_R$ is a strictly better against $\sigma_C$ in the learning game than $\sigma_R$ against $\sigma_C$. Thus, the example proves that if a learning heuristic is uncoupled, then it cannot both converge to Nash equilibrium in all games and be a Nash equilibrium learning heuristic of the learning game.

---

[6]Obviously, to fit the examples within our model outlined in the previous section, we would need to normalize payoffs of games in $\mathbf{U}^1$ and $\mathbf{U}^2$ to fit in $[0, 1]$ with affine transformations.

## 4  How General is the Counterexample?

Note that our arguments do not depend on other properties of learning rules that have been discussed in the literature such as $m$-periods of recall, $m$-memory, stationarity, stochasticity (Hart and Mas-Colell, 2006) or radical uncoupledness (Foster and Young, 2006, Germano and Lugosi, 2007, Young, 2009). Sometimes some of those properties are imposed on top of uncoupledness to obtain positive or negative results on learning Nash equilibrium. Our observations hold for any such uncoupled learning heuristics converging to Nash equilibrium in all games.

It has been argued that correlated equilibrium (Aumann, 1974) has a better learning foundation than Nash equilibrium as there are simple learning heuristics approaching correlated equilibrium but not necessarily Nash equilibrium (Foster and Vohra, 1997, Fudenberg and Levine, 1999, and Hart and Mas-Colell, 2000, 2001, 2003). Note that in both games, $\mathbf{u}^1$ and $\mathbf{u}^2$, the unique correlated equilibrium is the unique Nash equilibrium. Similarly, in both games the unique Nash equilibrium coincides with the sets of iterated admissible strategies, rationalizable strategies (Bernheim, 1984, Pearce, 1984), and minimal curb sets (Basu and Weibull, 1991). Thus, the counterexample shows more generally that if a learning heuristic is uncoupled, then it cannot both converge to correlated equilibrium (the set of iterated admissible strategy profiles, rationalizable strategy profiles, or minimal curb sets, respectively) in all finite games and be a Nash equilibrium learning heuristic of the learning game.

May be asking for convergence in *all* games is requiring too much. Is there a generic and economically interesting class of games for which we could obtain a possibility result? In particular, one may hope for a possibility result when restricting to classes of games in which learning is known to be well-behaved. Note that our arguments make use of two-player games only. In fact, we used $2 \times 2$ games only. Moreover, both games $\mathbf{u}^1$ and $\mathbf{u}^2$ are ordinal potential games (in the sense of Monderer and Shapley, 1996), with ordinal potentials given, respectively, by

$$P^1 = \begin{array}{c} \\ b \\ c \end{array} \begin{array}{c} a \quad b \\ \left( \begin{array}{cc} 5 & 8 \\ 10 & 9 \end{array} \right) \end{array} \qquad P^2 = \begin{array}{c} \\ b \\ c \end{array} \begin{array}{c} a \quad b \\ \left( \begin{array}{cc} 2 & 3 \\ 1 & -1 \end{array} \right) \end{array}$$

Note further that both games satisfy strategic complements. I.e., order actions by $a < b < c$. Then both stage-games satisfy increasing differences, $u_i(x'', y') - u_i(x', y') \leq u_i(x'', y'') - u_i(x', y'')$ for any actions $x'' > x'$, $y'' > y'$, a property that facilitates convergence of many learning heuristics (e.g., Milgrom and Roberts, 1990). Both games we used have a unique Nash equilibrium which is in pure actions. Thus, the observation is not due to the fact of converging to a "wrong" equilibrium or miscoordination. All games we used are generic and "non-pathological". Often stronger positive results on learning have been obtained for generic games only (see for instance Germano and Lugosi, 2007, Young, 2009).[7] Thus, one cannot hope to obtain positive results for those subclasses of games. This seems particularly frustrating for the class of $2 \times 2$ games, ordinal potential games, and games with strategic complements for which simple adaptive learning heuristics are known to converge to equilibrium, see Fudenberg and Levine (1998).

We can summarize our observations so far in the following proposition:

**Proposition 1** *Let* **U** *be the class of all finite games. If a learning heuristic is uncoupled, then it cannot both converge to Nash equilibrium in all games in* **U** *and be a Nash equilibrium learning heuristic of the learning game based on* **U**. *This holds even if we restrict* **U** *to be the class of all generic finite games, all two-player games, all $2 \times 2$ games, all games with pure Nash equilibrium, all games with a unique Nash equilibrium, all games with strategic complements, all ordinal potential games, or any union thereof, or if we weaken convergence to correlated equilibrium, the set of iterated admissible strategy profiles, rationalizable strategy profiles, or minimal curb sets.*

## 5 Incentives for Strategic Teaching

In this section, we are interested in establishing a lower bound on the "average" long run payoffs that can be achieved against an opponent who adopted an uncoupled learning heuristic that converges to Nash equilibrium in all games. This allows us to draw a parallel to reputation results in the repeated games literature.

---

[7]The mathematical notion of genericity does not always coincide with economic relevance in game theory. E.g., normal-form games associated to extensive-form games are typically non-generic. Yet, we focus here on learning of normal-form solution concepts. Moreover, the games used in our counterexample are not irrelevant in economics. In fact, game $\mathbf{u}^2$ can be viewed as a textbook-style Cournot duopoly in which player $i$'s profit function is $\max\{10.9 - q_i - q_{-i}, 0\} \cdot q_i - 0.1 \cdot q_i$, and for which we deleted all quantities $q_i$ and $q_{-i}$ except for the Cournot Nash equilibrium quantity as well as the Stackelberg leader and follower quantities of each player and suitably rounded payoffs.

Define player $i$'s pure best response correspondence for the stage-game $\mathbf{u} \in \mathbf{U}$ by

$$B_i(\mathbf{u})(a_{-i}) := \{a_i \in A_i : u_i(a_i, a_{-i}) \geq u_i(a'_i, a_{-i}) \text{ for all } a'_i \in A_i\}.$$

Further, define player $i$'s worst Stackelberg leader payoff in game $\mathbf{u} \in \mathbf{U}$ by

$$\ell_i(\mathbf{u}) := \max_{a_i \in A_i} \min_{a_{-i} \in B_{-i}(\mathbf{u})(a_i)} u_i(a_i, a_{-i}).$$

In this definition, the Stackelberg leader is pessimistic since in case of multiple best responses of the follower, he assumes that the follower chooses the best response that is worst to him. This seems appropriate since our aim is to establish a lower bound. Yet, best responses are unique in generic games. Thus, for "average" long run payoffs (i.e., "averaged" over all games with respect to the Lebesgue measure on the space of games), the "pessimistic" selection from the follower's best response correspondence does not matter.

Let

$$L_i(\mathbf{U}) := \int_{\mathbf{U}} \ell_i(\mathbf{u}) d\lambda$$

be the "average" of Stackelberg leader payoffs over games in the Lebesgue measurable class $\mathbf{U}$ with respect to the Lebesgue measure $\lambda$.

A player who faces an opponent following an uncoupled learning heuristic that converges to Nash equilibrium in all finite two-player games that possess a pure Nash equilibrium can guarantee herself almost surely at least the Stackelberg leader payoff averaged over all games with a suitable strategic teaching strategy. Moreover, this payoff is strictly larger than adopting the uncoupled learning heuristic as well and converging to a Nash equilibrium in all games. In this sense, there is a strict positive incentive for strategic teaching to prevent learning from converging to Nash equilibrium in all games. This is stated more formally in the following proposition:

**Proposition 2** *Let $\mathbf{U}$ be the class of finite two-player games that possess a pure Nash equilibrium. For any profile of uncoupled learning heuristic $(\sigma_R, \sigma_C)$ that converges to Nash equilibrium in all games in $\mathbf{U}$, there exists a learning heuristic $\tilde{\sigma}$ (i.e., a "strategic teaching heuristic") such that when Rowena follows $\tilde{\sigma}$ and Colin follows $\sigma_C$, we have*

$$V_R((\tilde{\sigma}, \sigma_C), \mathbf{U}) \geq L_R(\mathbf{U}) > V_R((\sigma_R, \sigma_C), \mathbf{U}).$$

*This holds even if we weaken convergence to the set of correlated equilibria, iterative admissible action profiles, rationalizable action profiles, or minimal curb sets.*

The proof is contained in Appendix A.1. We show that a player who is facing an opponent with an uncoupled learning heuristic converging to Nash equilibrium can guarantee herself with a "strategic teaching heuristic" at least the Stackelberg leader payoff in every generic two-player game that possesses a pure Nash equilibrium. This is because for every generic two-player game and Stackelberg outcome, there is a generic game in which this Stackelberg outcome is the unique Nash equilibrium. This game differs from the first in the player's payoffs only but not in the opponent's payoffs. The player can then "pretend" to be in this game whenever the original game is played and strategically teach the opponent to jointly reach almost surely the outcome in the second game because the opponent follows an uncoupled learning heuristic. In this sense, the proof of Proposition 2 generalizes our counterexample as it shows that there is an opportunity for strategic teaching in *every* generic two-player game that possesses a pure Nash equilibrium. It does not imply though that the strategic teacher earns a strict higher payoff than in equilibrium in every game. The fact that the "average" over Stackelberg leader payoffs is strictly better than the "average" limit-of-means payoffs of $\boldsymbol{\sigma}$ is perhaps not obvious. It is known that there are finite two-player games with a pure Nash equilibrium in which the Stackelberg leader payoff as defined here can be strictly below a Nash equilibrium payoff (e.g., Başar and Olsder, 1999, p. 132-133). But such games must be non-generic and won't get positive measure when "averaging" expected limit-of-means payoffs over all games in $\mathbf{U}$ with respect to $\lambda$. In generic games with pure equilibrium, the worst Stackelberg leader payoff is weakly above equilibrium payoff. Since our counterexample shows that there are also open subsets of games where the worst Stackelberg leader payoff is strictly above equilibrium payoff, the strict inequality in the proposition follows.

# 6 "Possibility" Results

Uncoupled learning heuristics depend only on the player's own payoff function and the behavior of the opponent but not directly on the opponent's payoffs. With such learning heuristics "play has a decentralized character, and no player can, alone, recognize a Nash equilibrium" (Hart and Mas-Colell, 2006, p. 287). Thus, it is reasonable to expect that we could obtain a positive result when we restrict to the class of games where just both player's rationality suffices to find the Nash equilibrium action. In Nash equilibrium of such games, each player plays as if he solves individual decision problems. Nothing about the opponent's payoff function needs to learned by the players in order to find the solution to the game. Hence, we may label such games "strategically trivial". We will show that indeed for such classes of games a possibility result can be obtained but that the possibility results fails when any measurable subset of games with positive Lebesque measure outside this class is considered as well. In this sense, a foundation of learning in games is possible as long as just "strategically trivial" games are considered.

We use as a rationality criterion the notion of admissibility, i.e., the avoidance of weakly dominated actions.

**Definition 4 (Weak Dominance)** *An action $a_i \in A$ is weakly dominated if there is a mixed action $\alpha_i \in \Delta(A)$ such that*

$$\begin{aligned} u_i(\alpha_i, a_{-i}) &\geq u_i(a_i, a_{-i}) \text{ for all } a_{-i} \in A, \text{ and} \\ u_i(\alpha_i, a_{-i}) &> u_i(a_i, a_{-i}) \text{ for some } a_{-i} \in A. \end{aligned}$$

*Let $D_i(u_i) \subseteq A$ denote the set of all actions that remain after deletion of all weakly dominated actions in a game in which player $i$'s utility function is $u_i$.*

*A game $\mathbf{u}$ is weak dominance solvable[8] (in one round) if for every player $i \in \{R, C\}$ and for all $a_i, a'_i \in D_i(u_i)$,*

$$u_i(a_i, a_{-i}) = u_i(a'_i, a_{-i}) \text{ for all } a_{-i} \in D_{-i}(u_{-i}).$$

*Denote by $\mathbf{WDS}$ the class of (one-round) weak dominance solvable games.*

Next we define a "richness" property of classes of games.

**Definition 5 (Product Class)** *A class of games $\mathbf{U}$ is a product class if $(u_i, u_{-i}), (\tilde{u}_i, \tilde{u}_{-i}) \in \mathbf{U}$ implies $(u_i, \tilde{u}_{-i}), (\tilde{u}_i, u_{-i}) \in \mathbf{U}$.*

Product classes of games are natural to consider in the context of uncoupled learning. Uncoupledness means that the learning heuristic cannot directly depend on opponents' payoffs. If a class of games is not a product class then the opponent's payoffs are not independent from the player's payoffs. Thus, an uncoupled learning heuristic could nevertheless condition implicitly on a subset of opponents' payoffs. Product classes of games are also motivated by our desire to find "maximal" classes of games for which we can obtain a positive result. A product class of games is closed under permutations of each player's payoff functions. Note that the class of (one-round) weak dominance solvable games is by definition a product class of games.

---

[8]There are various notions of "dominance solvability" in the literature. We follow Moulin (1979) except that we allow an action to be weakly dominated by a mixed action. Note that the definition does not necessarily imply that payoff functions are constant on *all* outcomes that remain after one round of elimination of weakly dominated actions for each player but just on outcomes that the player can unilaterally choose and remain after one round of weakly dominated actions.

**Proposition 3** *If $\mathbf{U}$ is a nonempty measurable class of games such that $\mathbf{U} \subseteq \mathbf{WDS}$, then there exists a learning heuristic that is uncoupled, converges to Nash equilibrium in all games in $\mathbf{U}$, and is a Nash equilibrium learning heuristic of the learning game based on $\mathbf{U}$. Conversely, if there is a measurable product class of games $\mathbf{U}$ with a pure Nash equilibrium such that $\mathbf{WDS} \subseteq \mathbf{U}$ and for which there exists a learning heuristic that is uncoupled, converges to Nash equilibrium in all games in $\mathbf{U}$, and is a Nash equilibrium learning heuristic of the learning game based on $\mathbf{U}$, then $\mathbf{U} \setminus \mathbf{WDS}$ must be of measure zero. This holds even if we weaken convergence to the set of correlated equilibria, iterative admissible action profiles, rationalizable action profiles, or minimal curb sets.*

The proof is contained in the Appendix A.2. The first part is straightforward. The proof of the converse is by contradiction. The idea is as follows: For any generic $\mathbf{u} = (u_i, u_{-i}) \notin \mathbf{WDS}$ with $\mathbf{u} \in \mathbf{U}$, we can find a generic game $\tilde{\mathbf{u}} = (\tilde{u}_i, \tilde{u}_{-i}) \in \mathbf{WDS}$ such that $(u_i, \tilde{u}_{-i}) \in \mathbf{U}$ and in which player $-i$ has an incentive to pretend being in $\mathbf{u}$ although the game might be $(u_i, \tilde{u}_{-i})$. Player $i$ with uncoupled learning behaves like in $\mathbf{u}$ giving $-i$ a higher payoff than in the unique Nash of $(u_i, \tilde{u}_{-i})$. Since these games are generic, they show up in the average long run payoff, which proves the converse.

In some sense, the "possibility" result is just another impossibility result as it fails when games beyond "strategically trivial" games are considered as well.

There are other classes of games for which we can obtain a possibility result. In any such a class a player's own payoff is sufficiently informative about a *particular* Nash equilibrium of the game as long as one restricts to the class. One such class is the class of common interest games studied in Aumann and Sorin (1995):

**Definition 6** *A game $\mathbf{u}$ is a* common interest *game if there is a profile of payoffs $(z_R, z_C) \in u_R(A) \times u_C(A)$ that strictly Pareto dominates all other profiles of payoffs, i.e.,*

$$z_R > w_R \text{ and } z_C > w_C$$

*for all $(w_R, w_C) \in u_R(A) \times u_C(A) \setminus \{(z_R, z_C)\}$, where $u_i(A)$ denotes the image of $u_i$.*

Let $\mathbf{CI}$ denote the class of common interest games.

Note that $(z_R, z_C)$ is unique in a common interest game. Yet, action profiles for which the payoff is $(z_R, z_C)$ may not be unique. Further note that every common interest game has at least one pure Nash equilibrium that yields the payoff profile $(z_R, z_C)$. Note further that the class of common interest games is in some sense less "strategically trivial" than one-round weak dominance solvable games as they may still involve coordination problems.

**Proposition 4** *If* **U** *is a nonempty measurable class of games such that* **U** $\subseteq$ **CI**, *then there exists a learning heuristic that is uncoupled, converges to Nash equilibrium in all games* **U**, *and is a Nash equilibrium learning heuristic in the learning game based on* **U**. *If* **U** *is a measurable product class of games with pure Nash equilibrium such that* **CI** $\subseteq$ **U**, *then there is no learning heuristic that is uncoupled, converges to Nash equilibrium in all games in* **U**, *and is a Nash equilibrium learning heuristic in the learning game based on* **U**. *This holds also if we weaken convergence to the set of correlated equilibria, iterative admissible action profiles, rationalizable action profiles, or minimal curb sets.*

The proof is contained in Appendix A.3. In a common interest game, if a player obtains his maximal stage-game payoff, then she has no incentive to reach another outcome through strategic teaching. For the first part of Proposition 4 we just need to show that there is an uncoupled learning heuristics leading to a Pareto efficient Nash equilibrium in every common interest game. This is done by modifying an uncoupled learning heuristic that has been used by Hart and Mas-Colell (2006) to show convergence of uncoupled learning in games with a pure Nash equilibrium such that it converges to efficient pure Nash equilibrium. The second part of Proposition 4 we show with a counterexample that is constructed with the help of product classes of games. Note that quite different from the class of one-round dominance solvable games, the class of common interest games is not a product class. (Thus, in the second part of Proposition 4 we could have written without loss of generality **CI** $\subsetneq$ **U**.) A player can infer from his payoff function and the fact that he faces only common interest games enough information about the opponent's payoffs so as to know a Nash equilibrium of the game. Since the second part of Proposition 4 holds for any product class of games with pure Nash equilibrium that contains common interest games, it holds also for the smallest such class that is generated by considering all permutations over all payoff functions in common interest games. In this sense, the second part of Proposition 4 shows an impossibility even if we only consider payoff function from common interest games.

# 7  Discussion

**Interpretation**

How to interpret our observations? First, we can interpret it from a *strategic* point of view. The fact that uncoupled Nash-converging learning heuristics are not Nash equilibrium in the learning game[9] means that a player has no incentive to stick to such a heuristic if opponents follow it.[10] Arguably such a strategic interpretation is meaningful and relevant only if players can be expected to "optimally choose" learning heuristics. This is not obvious since learning in games is studied precisely because players are not necessarily rational and may only learn to behave in the long run "as if" being rational. If players are unable to initially figure out how to behave optimally in some particular games then presumable they must be unable to choose optimally their learning heuristics in the arguably more complex learning game. Yet, there is experimental evidence that some subjects do indeed use sophisticated heuristics that involve strategic teaching of opponents who follow uncoupled learning heuristics. For instance, Camerer, Ho, and Chong (2002) estimate in experimental data learning models with adaptive experience-weighted attraction learning of players in repeated games but also sophisticated players who respond optimally to their forecasts of all others' behavior. A stark example is presented in Duersch, Kolb, Oechssler, and Schipper (2010), who report on an experiment in which human subjects played against a computer opponent programmed to myopic best response in a finitely repeated linear Cournot duopoly. Myopic best response is an uncoupled learning heuristic that is known to converge to Nash equilibrium in this game if both players follow it. Against the myopic best response computer, one subject played a surprising 4-cycle of quantities depicted by the upper (red) time series in Figure 1 and obtained a strictly larger average profit than the Stackelberg leader profit.[11] It turns out that this strategy is extremely close to the optimal strategy against myopic best response in this game (see Schipper, 2011, for the derivation of the optimal strategy). Similarly, Terracol and Vaksmann (2009) show in an experimental $3 \times 3$ Cournot duopoly that players may forego some immediate payoff in order to modify the opponent's future behavior. Moreover, players use strategic teaching to reach their favorable outcome. Hyndman, Ozbay, Schotter, and Ehrblatt (2012) show that subjects may use strategic teaching to help opponents converging to Nash equilibrium. Yet, their experimental set up involves two games in which the pure Nash equilibrium coincides with the Stackelberg outcome. This implies that the games lack an incentive to strategically teach an outcome that is not the pure Nash equilibrium. They conclude from their study and other studies in the literature "that what subjects attempt to teach is context dependent, and that their willingness to teach depends on the incentives to do so." To sum up, there is experimental evidence that some subjects are able to engage in strategic teaching in order to manipulate their opponents.

---

[9]Note that the impossibility result is not driven by the "rareness" of Nash equilibrium of the learning game. Quite to to contrary, Folk theorems of repeated games show that there are plenty of Nash equilibria of the learning game. The irony is that despite their abundance, none of them fits all requirements.

[10]Our notion of Nash equilibrium of the learning game is similar to Nash equilibrium of boundedly rational rules for playing normal-form games in Germano (2007). He studies the dynamic evolution of rules to play games, where a player's rule simply assigns a mixed action to each game. Recently we also learned about a closely related notion of "learning equilibrium" in computer science by Branfman and Tennenholtz (2004). They consider Nash equilibrium of learning heuristics but require Nash equilibrium game-by-game while we consider Nash equilibrium of learning heuristics "averaged" over classes of games. Monderer and Tennenholtz (2007) argue that learning equilibrium is the appropriate generalization of learning to optimize in the multi-agent context. Note that our observations apply also when using their stronger notion of "learning equilibrium".

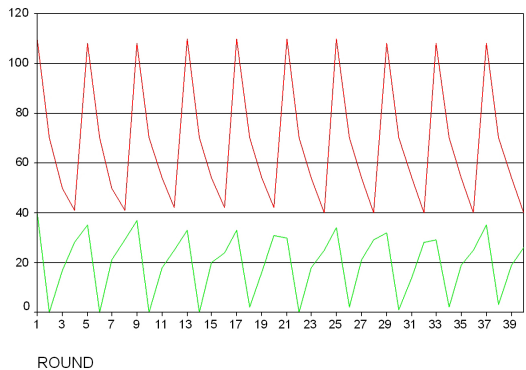

Figure 1: Almost optimal cycle played by a subject

A second interpretation can be provided in an *evolutionary* context. Instead of players choosing learning heuristics, the emergence of learning heuristics may be the result of mutations and evolutionary selection. Suppose there is a large population of players who are randomly matched to play two-player games drawn at random from a class of games $U$. For each game, each player has equal chance to play in the position of the row or column player (which symmetrizes games). Players are programmed to learning heuristics from a finite set. Each player's fitness (i.e., offsprings) is measured by the average long run payoffs over those games. (That is, evolution is assumed to be slower than it takes for average long run payoffs from learning to emerge.) Suppose now that initially the population is programmed to an uncoupled learning heuristic converging to Nash equilibrium in all games. Would such a population be robust to a small fraction of mutants that may invade with another heuristic to play in those games? The notion of evolutionary stability is intricate in repeated games. Yet, we believe that a reasonable notion of evolutionary stability in repeated games will require Nash equilibrium of the repeated game as a necessary condition (see for instance Binmore and Samuelson, 1992, Demichelis, 2013, Fudenberg and Maskin, 1990, Kim, 1994). Our negative observation on Nash equilibrium of the learning game does not force us to take a stand on the "right" notion of evolutionary stability in learning games as long as it requires Nash equilibrium. Under this provision, we conclude from our observations that if an uncoupled learning heuristics converges to Nash equilibrium in all games, then it cannot be evolutionary stable in the learning game.

---

[11]The payoff function is $u(a_i, a_{-i}) = \max\{109 - a_i - a_{-i}, 0\} \cdot a_i - a_i$. The Stackelberg leader quantity is 54, the Stackelberg following quantity is 27, and the Nash equilibrium quantity is 36.

A third interpretation is based on the idea that a learning heuristic when applied to the learning game should *learn itself*. One way of explaining the use of a particular learning heuristic is to assume that players could learn to learn. That is, they may not just use a learning heuristic to learn in the underlying games but since it should apply to *all* games, they may also use it to (meta-)learn the learning heuristic in a game where they could choose learning heuristics and payoffs are average long run payoffs. But this would just push the problem one level further. How do players find such a learning heuristic? Presumably there is a (meta-meta-)learning heuristic to (meta-)learn the learning heuristic to play in the games. Clearly, we are headed for an infinite regress.[12] There is no need for us to formalize this infinite regress problem here because our counterexample shows that there is already a problem at second level. There is no uncoupled learning heuristic converging to Nash equilibrium in all finite games that could learn itself. This is in contrast to the existence of universal learning heuristics for "single-person decision problems" in artificial intelligence (Schmidhuber, 2003).

**Mixed Equilibrium and Strictly Competitive Games**

We considered learning heuristics that converge to pure Nash equilibrium of the stage-game if such an equilibrium exists. Are our negative observations an artefact of focusing on pure equilibrium? Would it be possible to find uncoupled learning heuristics that converge to non-degenerate mixed equilibrium of the stage-game (in games that do have such equilibria) and that each player has an incentive to adopt if the opponent adopt it as well? Moreover, is there a possibility result when restricting to strictly competitive games, a class of games in which learning is usually "nice"? Note that when confining the analysis to strictly competitive games, the uncoupledness assumption of learning heuristics loses to a large extent its restrictiveness since each player's payoff function is informative about the opponent's payoff function. We will argue that restricting the previous analysis to pure equilibrium was for conceptual clarity and mathematical simplicity and that counterexamples can also be obtained within the class of strictly competitive games. Consider as a counterexample the "matching pennies" game

$$\begin{array}{c} & h & t \\ h & \begin{pmatrix} 1,-1 & -1,1 \\ -1,1 & 1,1 \end{pmatrix} \end{array}$$

---

[12]Similar but different infinite regress problems of how to decide how to decide ... have been studied by Mongin and Walliser (1988) and Lipman (1991).

The unique Nash equilibrium is the profile of non-degenerate mixed actions, $((\frac{1}{2},\frac{1}{2}),(\frac{1}{2},\frac{1}{2}))$. Now consider a "biased matching pennies" game,

$$\begin{array}{c} & h & t \\ h & \begin{pmatrix} 2,-1 & -1,1 \\ -1,1 & 1,1 \end{pmatrix} \end{array}$$

that just differs from the previous game in that Rowena's payoff from $(h,h)$ increased to 2. Colin's payoffs remain unchanged. Since Rowena's equilibrium mixed action must make Colin indifferent among his actions, Rowena's equilibrium mixed action remains unchanged. Colin's equilibrium action changes to $(\frac{2}{5},\frac{3}{5})$. He should put less weight on pure action $h$ so as to keep Rowena indifferent among her actions. Since Colin uses an uncoupled learning heuristic, his behavior in both games must be the same unless Rowena "communicates" her change of payoff through her different play. Note, however, that Rowena does not have an incentive to do so. If she "communicates" her biased payoff and both players use learning heuristics leading to Nash equilibrium (let's say in terms of almost-sure convergence of per-period behavior à la Hart and Mas-Colell, 2006, Theorem 7) in both games, then her long run payoff is 0.2. If instead she behaves in the biased matching pennies game like her learning heuristic in the matching pennies game, then Colin won't be able to learn about her biased payoffs and her long run payoff is 0.25. Since both games are generic, this simple example shows that we can apply arguments similar to the previous sections also to the case of convergence to non-degenerate mixed equilibrium and the class of strictly competitive games.

**How about Rational Learning?**

Kalai and Lehrer (1993) studied rational learners who know their own payoff matrix, don't know opponents' payoff matrices but are able to form probabilistic beliefs about opponents' strategies in infinitely repeated games, update beliefs according to Bayes rule, and chose strategies so as to maximize discounted expected utility. They show that rational learners must eventually play like a Nash equilibrium of the *repeated game*. As a corollary, they show that rational learners play like a Bayesian Nash equilibrium of the repeated game with uncertainty about opponents' payoff matrices. Moreover, they will eventually play like a Nash equilibrium of the real repeated game as if they know all players' payoffs.

If rational learners are myopic (as in the case of Jordan, 1991), i.e., they almost completely discount future payoffs, then our counterexample applies verbatim since any Nash equilibrium of the repeated game must become like repeated Nash equilibrium of the stage game. Thus, myopic rational learners converge to stage-game Nash equilibrium. Our counterexample shows that in many classes of games players do not have an incentive to adopt myopic rational learning if opponents adopt it.

If rational learners become extremely patient, then our counterexample slightly qualifies the results by Kalai and Lehrer (1993). Recall that they show that rational learning eventually plays like *a Nash equilibrium of the repeated game*. We know from Folk theorems that the set of repeated game Nash equilibria is typically large. It also contains the repeated stage-game Nash equilibrium. Our counterexample implies that patient rational learners won't play like a Nash equilibrium of the repeated game that consists of playing a stage-game Nash equilibrium from some stage onwards. That is, in games like our counterexample, *patient rational learning cannot converge to any repeated game Nash equilibrium but must work as an equilibrium selection device in repeated games*. To see this, consider our counterexample as a repeated game with one-sided incomplete information and known own payoff matrices. At the beginning of the repeated game, either payoff matrix $u^1$ or $u^2$ is drawn according to some non-degenerate probability distribution $(p, 1-p)$. Rowena is informed about the draw while Colin is kept ignorant. Note that Colin's payoff matrix in game $u^1$ is identical to his payoff matrix in game $u^2$. Thus, he has complete information about his own payoff matrix but incomplete information about his opponent's payoff matrix. According to a characterization result by Shalev (1994), every Bayesian Nash equilibrium of this repeated game is payoff-equivalent to a fully revealing Nash equilibrium of the repeated game.[13] Figure 2 shows the payoffs for the repeated games $u^1$ (left figure) and $u^2$ (right figure). The area shaped by the intermitted line indicates feasible payoffs of the repeated game with complete information. The thick vertical and horizontal solid lines mark the minimax payoffs of Rowena and Colin in those games, respectively. Consequently, the grey-shaded areas show equilibrium payoffs in the repeated game with complete information. The red dot indicates the stage-game Nash equilibrium payoff. There is no Bayesian Nash equilibrium of the repeated game with one-sided incomplete information in which players play like the repeated stage-game Nash equilibrium in $u^1$ when $u^1$ is drawn and the repeated stage-game Nash equilibrium in $u^2$ when $u^2$ is drawn. A necessary condition of Shalev (1994)'s characterization result is *incentive compatibility* for Rowena. In our setting this means that Rowena's payoff from playing $u^2$ must be weakly larger than her payoff in $u^2$ when playing like in $u^1$. Clearly, this is violated in our counterexample as she receives 13 in the stage-game Nash equilibrium of $u^2$ but 15 when she plays in $u^2$ like in (the stage-game Nash equilibrium of) $u^1$. Suppose now that rational learners would converge to the repeated stage-game equilibrium. Then previous arguments show that the rational teacher has an incentive to take advantage of such a learner. This would lead the play away from the repeated stage-game equilibrium, a contradiction. In fact, our strategic teaching strategy of playing in $u^2$ like in $u^1$ is the *best* strategy against Colin. This follows from a characterization result by Israeli (1999) in the repeated games literature (see Cripps and Thomas, 2003, for the case with discounting). Applied to our context, his result means that the best Rowena can do again Colin in $u^2$ is to play as if she plays in the zero-sum game in which her payoffs is the negative of Colin's payoffs. Note that this zero-sum game has the same pure best response structure as $u^1$. Rowena's stage-game Nash equilibrium strategy in $u^1$ minimizes Colin's payoff in $\mathbf{u}^2$. Thus, if she plays $u^2$ like $u^1$, then she plays like in the zero-sum game.[14]

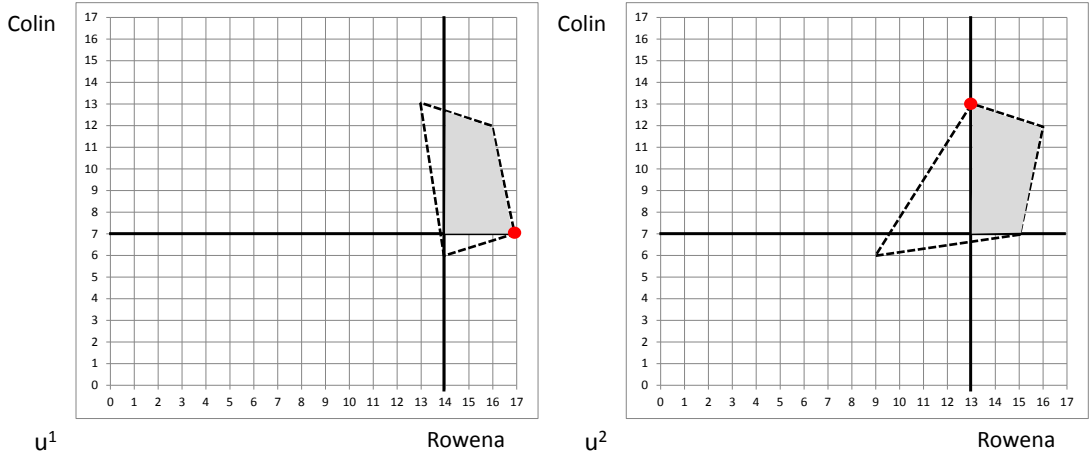

Figure 2: Nash Equilibrium Payoffs of the Repeated Games in the Counterexample

---

[13]There is one caveat. Kalai and Lehrer (1993) use discounting. Yet, it is known from Cripps and Thomas (2003) that the case with discounting but patient players is very similar to Shalev (1994)'s result. See Sorin (1999) for a synthesis.

To sum up, our counterexample shows that patient rational learning of Kalai and Lehrer (1993) selects among Nash equilibria of the repeated game. In particular, it can not converge to play like repeated stage-game Nash equilibrium in every game. The fact that patient rational learning must select among Nash equilibria of some repeated games matches with the known fact in the repeated games literature that small uncertainty over repeated games may shrink the set of repeated game equilibrium payoffs dramatically (see Shalev, 1994, for an example attributed to Robert Aumann). Rational learning does not eliminate opportunities for strategic teaching as already observed by Kalai and Lehrer (1993, Example 2.4). In fact, the patience of rational learners facilitates strategic teaching of opponents.

**Beyond Uncoupledness**

Our observations depend heavily on the prominent assumption in the literature that learning heuristics are uncoupled. One may conjecture that positive results can be obtained when simple coupled learning heuristics are allowed. This conjecture cannot be true however. Our observations show already that negative results can obtained even if uncoupledness is slightly weakened. This follows from our observation that negative results hold even if we restrict to subclasses of games such as potential games or games with strategic complementarities. In such cases, even uncoupled learners are allowed to to make use to some extent of opponents' payoff functions because any of the opponents' payoff function must be consistent with the subclass of games. It also follows from our discussion of rational learning. Although at the first glance rational learning à la Kalai and Lehrer (1993) appears to be uncoupled in the sense that it does not depend on the exact payoff function of opponents, it still makes use of *some* information about opponents' payoffs via the absolute continuity assumption. Each player's prior belief is not allowed to assign zero probability to strategies of opponents that are actually used. Since players are rational learners, their strategies actually used must be optimal with respect to their payoff functions. Thus, rational learning depends on some information about opponents' payoff functions via the absolute continuity assumption on the prior. Hence, our observations apply also to learning heuristics that to some extent make use of *some* information about opponents' payoffs.[15]

---

[14]This does not imply that our Proposition 2 necessarily extends to patient rational learners. First, strategic teaching opportunities depend on the prior. Rowena is able to strategically teach Colin because Colin faces the particular uncertainty over Rowena's payoffs matrices. Second, patient rational learners won't necessarily learn to play a Stackelberg follower action. Rather, strategic teachers might be constrained by the minimax actions of patient rational learners. This follows from Israeli (1999) and Cripps and Thomas (1995) for the case without discounting and from Cripps, Schmidt, and Thomas (1996) for the case with discounting but patient players.

So the challenge is on how exactly to go beyond uncoupleness of learning heuristics in a suitable way such that positive results can be obtained and learning heuristics remain "simple". Although uncoupledness has been motivated by Sergiu Hart as an informational requirement of simple and efficient learning heuristics,[16] we doubt that it formalizes appropriately a *necessary* property of a notion of "simple" learning heuristics. For instance, "imitate-the-best" (Vega-Redondo, 1997) is an extremely simple learning heuristic that is clearly coupled. Yet, it also does not necessarily lead to Nash equilibrium. But it is surprisingly robust to manipulation attempts by sophisticated opponents (see Duersch, Oechssler, and Schipper, 2012, 2014). Moreover, Schipper (2009) shows that noisy imitate-the-best learning can manipulate uncoupled learning heuristics such as myopic best response in some economically relevant classes of games (that include our counterexample), which demonstrates that strategic teaching of uncoupled learners does not necessarily require sophisticated strategic behavior on part of the teacher. Finding general notions of a "simple" learning heuristics that are also immune to manipulation by strategic teaching is a challenge left for further research. On a practical level, such learning heuristics and the knowledge of their long run properties may be useful for implementation problems in which we may want to allow agents to learn but want to avoid manipulation of learners. On a conceptual level, it would give us confidence in the wide-spread use of Nash equilibrium as solution concept for economic models.

---

[15] Our counterexample applies also to strategic teaching heuristics that are uncoupled. Note that in this counterexample the strategic teaching heuristic is constant in the opponents' payoff functions (within the union of open neighborhoods of games) and thus uncoupled. Requiring strategic teaching heuristics to be uncoupled may not allow for general impossibility results though. Yet, our observations suggest that there is an economic argument against uncoupleness. A strategic teacher may gain by obtaining information about opponents' payoffs and thus be willing to pay for it.

[16] Presidential Address at The Third World Congress of the Game Theory Society (GAMES 2008). See http://www.ma.huji.ac.il/hart/abs/dyn-p.html.

# A Proofs

## A.1 Proof of Proposition 2

Fix a generic stage-game $\mathbf{u} = (u_i, u_{-i}) \in \mathbf{U}$. Note that in generic games best responses are unique. Hence, we can write

$$a_i^L := \arg\max_{a_i \in A} u_i(a_i, a_{-i}) \text{ s.t. } a_{-i} = B_{-i}(\mathbf{u})(a_i)$$

and

$$a_{-i}^F := B_{-i}(\mathbf{u})(a_i^L)$$

for the Stackelberg leader and follower actions of the stage-game $\mathbf{u}$, respectively.

Since $\mathbf{U}$ is the class of finite two-player games that possess a pure Nash equilibrium, there exists a game $(\tilde{u}_i, u_{-i}) \in \mathbf{U}$ such that

(a) $(a_i^L, a_{-i}^F)$ is the unique Nash equilibrium of $(\tilde{u}_i, u_{-i})$, and

(b) $\tilde{u}_i(a_i^L, a_{-i}^F) \geq \ell_i(\tilde{u}_i, u_{-i})$.

E.g., let $\tilde{u}_i$ be such that

(A) $a_i^L$ strictly dominates all other actions in $A$ for player $i$, i.e., for any $\alpha_i \in \Delta(A)$ such that $\alpha_i(a_i^L) < 1$, we have $\tilde{u}_i(a_i^L, a_{-i}) > \tilde{u}_i(\alpha_i, a_{-i})$ for any $a_{-i} \in A$,

(B) player $i$'s payoff from the profile $(a_i^L, a_{-i}^F)$ strictly dominates any other payoff, i.e., $\tilde{u}_i(a_i^L, a_{-i}^F) > \tilde{u}_i(\mathbf{a})$ for any $\mathbf{a} \in A \times A$, $\mathbf{a} \neq (a_i^L, a_{-i}^F)$.

(A) implies (a) since $(\tilde{u}_i, u_{-i})$ is generic, $a_i^L$ is the unique best response of player $i$ to $a_{-i}^F$ in $(\tilde{u}_i, u_{-i})$ and $a_{-i}^F = B_{-i}(\mathbf{u})(a_i^L) = B_{-i}(\tilde{u}_i, u_{-i})(a_i^L)$. (B) implies (b). Moreover, both (A) and (B) are consistent in the sense that for any generic $\mathbf{u} \in \mathbf{U}$ there exist $(\tilde{u}_i, u_{-i}) \in \mathbf{U}$ such that both (A) and (B) hold.

Since $(\sigma_i, \sigma_{-i})$ is a profile of uncoupled learning heuristics that converges almost surely to the stage-game Nash equilibrium in all games in $\mathbf{U}$, it must almost surely converge to the Nash equilibrium $(a_i^L, a_{-i}^F)$ in the game $(\tilde{u}_i, u_{-i})$. Let player $i$ now follow a learning heuristic $\tilde{\sigma}_i$ that in both games, $\mathbf{u}$ and $(\tilde{u}_i, u_{-i})$, behaves like $\sigma_i$ in game $(\tilde{u}_i, u_{-i})$. (Thus, when in $\mathbf{u}$, player $i$ "pretends" to be in $(\tilde{u}_i, u_{-i})$.) Since $\sigma_{-i}$ is an uncoupled learning heuristic, player $-i$ behaves almost surely identically in $\mathbf{u}$ and $(\tilde{u}_i, u_{-i})$ when player $i$ follows $\tilde{\sigma}_i$. Thus, in both games, $\mathbf{u}$ and $(\tilde{u}_i, u_{-i})$, almost every play path consists of $(a_i^L, a_{-i}^F)$ being played from some point on. Hence,

$$\liminf_{T \to \infty} \mathbb{E}_{(\tilde{\sigma}_i(\mathbf{u}), \sigma_{-i}(\mathbf{u}))} \left[ \frac{1}{T} \sum_{t=1}^{T} u_i(\mathbf{a}^t(\tilde{\sigma}_i(\mathbf{u}), \sigma_{-i}(\mathbf{u}))) \right] \geq \ell_i(\mathbf{u})$$

and

$$\liminf_{T \to \infty} \mathbb{E}_{(\tilde{\sigma}_i(\tilde{u}_i, u_{-i}), \sigma_{-i}(\tilde{u}_i, u_{-i}))} \left[ \frac{1}{T} \sum_{t=1}^{T} \tilde{u}_i(\mathbf{a}^t(\tilde{\sigma}_i(\tilde{u}_i, u_{-i}), \sigma_{-i}(\tilde{u}_i, u_{-i}))) \right] \geq \ell_i(\tilde{u}_i, u_{-i}).$$

Since the argument holds for almost any repeated stage-game $\mathbf{u} \in \mathbf{U}$, we must have

$$V_i((\tilde{\sigma}_i, \sigma_{-i}), \mathbf{U}) \geq L_i(\mathbf{U}).$$

It is known that there are finite two-player games with a pure Nash equilibrium in which the (worst) Stackelberg leader payoff as defined here can be strictly below a Nash equilibrium payoff (e.g., Başar and Olsder, 1999, p. 132-133). But such games must be non-generic. In a generic game, best responses are unique. Thus, in generic games a Stackelberg leader can guarantee herself at least her best Nash equilibrium payoff because if the Stackelberg leader chooses the action corresponding to her most preferred pure Nash equilibrium, then the opponent best responds uniquely with his corresponding Nash equilibrium action. When "averaging" limit-expected-mean payoffs over all games in $\mathbf{U}$ with a Lebesgue measure, non-generic games must have measure zero. Thus $L_i(\mathbf{U}) \geq V_i((\sigma_i, \sigma_{-i}), \mathbf{U})$.

In our example $\mathbf{u}^2$ from Section 3 we observe that the Stackelberg leader achieves a payoff that is strictly larger than in the unique Nash equilibrium. Since $\mathbf{u}^2$ is generic, this holds for an open neighborhood $\mathbf{U}^2 \subseteq \mathbf{U}$ of the stage-game. Since any nonempty open neighborhood must have strict positive Lebesgue measure, we must have $L_i(\mathbf{U}) > V_i((\sigma_i, \sigma_{-i}), \mathbf{U})$.

Finally, note that in above arguments, $(a_i^L, a_{-i}^F)$ is the unique and strict Nash equilibrium of $(\tilde{u}_i, u_{-i})$ in which player $i$ plays a strict dominant action. Thus, it is also the correlated equilibrium, iterative admissible action profile, rationalizable action profile, and minimal curb set of $(\tilde{u}_i, u_{-i})$. Hence, the result holds also for uncoupled learning heuristics that converge to correlated equilibrium, iterative admissible action profiles, rationalizable action profiles or minimal curb sets, respectively. This completes the proof of the proposition. □

## A.2 Proof of Proposition 3

Let $\mathbf{U} \subseteq \mathbf{WDS}$ with $\mathbf{u} \in \mathbf{U}$. Any action that remains after one round of elimination of weakly dominated actions in the stage-game $\mathbf{u}$ is a Nash equilibrium action. For each player $i$, select $\sigma_i$ that in every stage-game $\mathbf{u} \in \mathbf{U}$ chooses an action $a_i \in D_i(u_i)$ for every history. Then the profile $\boldsymbol{\sigma} = (\sigma_i, \sigma_{-i})$ selects a Nash equilibrium in every stage-game $\mathbf{u} \in \mathbf{U}$. Require also that the action that $\sigma_i$ selects for player $i$ in $(u_i, \tilde{u}_{-i}) \in \mathbf{U}$ is identical to the action that it selects in $\mathbf{u} \in \mathbf{U}$, for every $\tilde{u}_{-i}$. Then $\sigma_i$ is uncoupled. Moreover, it follows that the profile of such learning heuristics is a Nash equilibrium in the learning game based on $\mathbf{U}$.

We prove the converse by contradiction. Let $\mathbf{WDS} \subseteq \mathbf{U}$ and let $\mathbf{u} \in \mathbf{U}$ be a generic stage-game with $\mathbf{u} \notin \mathbf{WDS}$. Fix a history of play of $\mathbf{u}$ that emerges from the profile of uncoupled learning heuristics $\boldsymbol{\sigma} = (\sigma_i, \sigma_{-i})$ that converges to Nash equilibrium in all games in $\boldsymbol{U}$. Since every game in $\boldsymbol{U}$ has a pure Nash equilibrium by definition, there is a Nash equilibrium of the stage-game $\mathbf{u}$ to which $\boldsymbol{\sigma}(\mathbf{u})$ converges almost surely. Denote it by $\boldsymbol{a}^\infty(\boldsymbol{\sigma}(\mathbf{u})) = (a_i^\infty(\boldsymbol{\sigma}(\mathbf{u})), a_{-i}^\infty(\boldsymbol{\sigma}(\mathbf{u})))$.

We claim that there exists a generic stage-game $\tilde{\mathbf{u}} = (\tilde{u}_i, \tilde{u}_{-i}) \in \mathbf{WDS}$ such that

(0) $(u_i, \tilde{u}_{-i})$ is a generic game in $\mathbf{U}$,

(i) any Nash equilibrium action of player $i$ in the stage-game $(u_i, \tilde{u}_{-i})$ differs from $a_i^\infty(\boldsymbol{\sigma}(\mathbf{u}))$,

(ii) $\tilde{u}_{-i}(\boldsymbol{a}^\infty(\boldsymbol{\sigma}(\mathbf{u}))) > \tilde{u}_{-i}(\boldsymbol{a}^*)$ for any Nash equilibrium $\boldsymbol{a}^*$ of the stage-game $(u_i, \tilde{u}_{-i})$.

Intuitively, (ii) implies that player $-i$ has an incentive to pretend being in game $\mathbf{u}$ even though the stage-game is $(u_i, \tilde{u}_{-i})$. Together with (i), player $-i$ has an incentive to not let the play converge to a Nash equilibrium of $(u_i, \tilde{u}_{-i})$. (0) implies that this is relevant for the learning game based on $\mathbf{U}$.

(0) follows because $\tilde{\mathbf{u}} \in \mathbf{WDS} \subseteq \mathbf{U}$, both $\tilde{\mathbf{u}}$ and $\mathbf{u}$ are generic, $\mathbf{U}$ is a product class of finite two-player games, and $(u_i, \tilde{u}_{-i})$ has a pure Nash equilibrium. To see the last point, pick for player $-i$ an action $a_{-i} \in D_{-i}(\tilde{u}_{-i})$ and note that because $\tilde{\mathbf{u}}$ is generic there must be a pure and unique best response $a_i$ to it by player $i$. Action $a_{-i}$ is the unique best response to $a_i$ in $(u_i, \tilde{u}_{-i})$ because $\tilde{\mathbf{u}} \in \mathbf{WDS}$ and $\tilde{\mathbf{u}}$ is generic. Thus, $(a_i, a_{-i})$ is a strict pure Nash equilibrium of $(u_i, \tilde{u}_{-i})$.

(i) follows from $\mathbf{u} \notin \mathbf{WDS}$. To see this, note that $\mathbf{u} \notin \mathbf{WDS}$ implies that for some

player $i$ there exist $a_i^{**}, a_i^* \in D_i(u_i)$ such that

$$u_i(a_i^{**}, a_{-i}) > u_i(a_i^*, a_{-i}) \text{ for some } a_{-i} \in D_{-i}(u_{-i}).$$

If in addition

$$u_i(a_i^{**}, a_{-i}) \geq u_i(a_i^*, a_{-i}) \text{ for all } a_{-i} \in A,$$

then $a_i^{**}$ weakly dominates $a_i^*$, a contradiction to $a_i^* \in D_i(u_i)$. (In the last inequality, we could have written ">" since the game is generic.) Thus, we must also have

$$u_i(a_i^{**}, a_{-i}) < u_i(a_i^*, a_{-i}) \text{ for some } a_{-i} \in A. \tag{3}$$

Note since $\mathbf{u}$ is generic, there exist $a_{-i}^{**}, a_{-i}^* \in A$ such that $a_i^{**}$ is the unique best response to $a_{-i}^{**}$ and $a_i^*$ is the unique best response to $a_{-i}^*$.

Since $\boldsymbol{\sigma}$ converges to a pure Nash equilibrium of $\mathbf{u}$, we must have $a_i^\infty(\boldsymbol{\sigma}(\mathbf{u})) \neq a_i^{**}$ or $a_i^\infty(\boldsymbol{\sigma}(\mathbf{u})) \neq a_i^*$. With loss of generality, assume the latter case (otherwise replace * with ** in below arguments), $a_i^\infty(\boldsymbol{\sigma}(\mathbf{u})) \neq a_i^*$. We can choose a generic stage-game $\tilde{\mathbf{u}} \in \mathbf{WDS}$ such that $a_{-i}^*$ strictly dominates all other actions in $A$ for player $-i$, i.e., for all $a_{-i} \in A \setminus \{a_{-i}^*\}$,

$$\tilde{u}_{-i}(a_i, a_{-i}^*) > \tilde{u}_{-i}(a_i, a_{-i}) \text{ for all } a_i \in A.$$

Then by construction, $a_{-i}^*$ is the unique Nash equilibrium action of player $-i$ in the game $(u_i, \tilde{u}_{-i})$. Hence, $\mathbf{a}^* = (a_i^*, a_{-i}^*)$ is the unique Nash equilibrium of $(u_i, \tilde{u}_{-i})$. It follows that player $i$'s Nash equilibrium action in the game $(u_i, \tilde{u}_{-i})$ is different from $a_i^\infty(\boldsymbol{\sigma}(\mathbf{u}))$, which finishes the proof of (i).

To prove (ii), note that by previous arguments it is sufficient to show

$$\tilde{u}_{-i}(\boldsymbol{a}^\infty(\boldsymbol{\sigma}(\mathbf{u}))) > \tilde{u}_{-i}(\mathbf{a}^*),$$

for the unique Nash equilibrium $\mathbf{a}^*$ of the game $(u_i, \tilde{u}_{-i})$.

Note that $a_{-i}^\infty(\boldsymbol{\sigma}(\mathbf{u})) \neq a_{-i}^*$. Suppose not, then since $a_i^*$ is the unique best response to $a_{-i}^*$ in the stage-game $\mathbf{u}$, it follows that $a_i^\infty(\boldsymbol{\sigma}(\mathbf{u})) = a_i^*$, a contradiction to the assumption above that $a_i^\infty(\boldsymbol{\sigma}(\mathbf{u})) \neq a_i^*$.

We can choose $\tilde{u}_{-i}$ such that

$$\tilde{u}_{-i}(a_i^\infty(\boldsymbol{\sigma}(\mathbf{u})), a_{-i}^*) = \tilde{u}_{-i}(\boldsymbol{a}^\infty(\boldsymbol{\sigma}(\mathbf{u}))) + \varepsilon = \tilde{u}_{-i}(\mathbf{a}^*) + 2\varepsilon$$

for some $\varepsilon > 0$. This makes $\tilde{u}_{-i}(\boldsymbol{a}^{\infty}(\boldsymbol{\sigma}(\mathbf{u})))$ sufficiently large in order to satisfy

$$\tilde{u}_{-i}(\boldsymbol{a}^{\infty}(\boldsymbol{\sigma}(\mathbf{u}))) \;>\; \tilde{u}_{-i}(\mathbf{a}^*),$$

while continuing to satisfy

$$\tilde{u}_{-i}(a_i^{\infty}(\boldsymbol{\sigma}(\mathbf{u})), a_{-i}^*) \;>\; \tilde{u}_{-i}(\boldsymbol{a}^{\infty}(\boldsymbol{\sigma}(\mathbf{u}))),$$

a necessary condition for $a_{-i}^*$ being strict dominant. This finishes the proof of (ii). Note that (i) and (ii) (and (0)) can be satisfied simultaneously.

Note that if player $i$ adopts $\sigma_i$, then player $-i$ can strictly improve her long run payoff in the repeated stage-game $(u_i, \tilde{u}_{-i})$ with $\sigma_{-i}^*$ that satisfies $\sigma_{-i}^*(\mathbf{u}) = \sigma_{-i}^*(u_i, \tilde{u}_{-i}) = \sigma_{-i}(\mathbf{u})$. That is, in game $(u_i, \tilde{u}_{-i})$, player $-i$ pretends to be in $\mathbf{u}$. Since both stage-games $\mathbf{u}$ and $(u_i, \tilde{u}_{-i})$ are generic, our arguments above hold also for open neighborhoods of them. Since any nonempty open neighborhoods must have strict positive Lebesgue measure, player $-i$'s "average" long-run payoff from $\sigma_{-i}^*$ in the learning game when player $i$ follows $\sigma_i$ is strictly larger than from $\sigma_{-i}$. Thus, $\boldsymbol{\sigma}$ is not a Nash equilibrium of the learning game, a contradiction.

In the above arguments, $(u_i, \tilde{u}_{-i})$ has a unique and strict Nash equilibrium in which one player plays a strict dominant action. Thus, it must also be the correlated equilibrium, the iterative admissible action profile, the rationalizable action profile, and the minimal curb set. Thus, analogous arguments apply when we weaken convergence to the set of correlated equilibria, iterative admissible action profiles, rationalizable action profiles, or minimal curb sets, respectively. $\square$

## A.3 Proof of Proposition 4

We show that there exists an uncoupled learning heuristic that in every game in **CI** leads to a Nash equilibrium with a payoff profile that strictly Pareto dominates any other payoff profiles. We do this by slightly modifying a learning heuristic used in Hart and Mas-Colell (2006, Proof of Theorem 3). Consider the following learning heuristic: for any $t \geq 2$,

- If $(a_R^{t-2}), a_C^{t-2}) = (a_R^{t-1}, a_C^{t-1})$, $a_i^{t-1}$ is a best response to $a_{-i}^{t-1}$, and the payoff obtained by player $i$ in $t-2$ and $t-1$ is player $i$'s maximal payoff of the stage-game, then player $i$ plays $a_i^t = a_i^{t-1}$.

- Otherwise, player $i$ randomizes uniformly in $t$ over all actions.

This learning heuristics is uncoupled; player $i$ does not condition on the opponent's payoffs. It is easy to see that if both players follow the heuristic in any game in **CI**, then they will reach a pure Nash equilibrium corresponding to the strict Pareto dominant payoff profile almost surely. The arguments are analogous to Hart and Mas-Colell (2006, Proof of Theorem 3). This proves the first part of Proposition 4.

Let **U** be a product class of games such that each stage-game in **U** has a pure Nash equilibrium and $\mathbf{CI} \subseteq \mathbf{U}$. Suppose by contradiction that there exists a profile of uncoupled learning heuristics $\boldsymbol{\sigma} = (\sigma_R, \sigma_C)$ that converges to Nash equilibrium in all games in **U** and is a Nash equilibrium learning heuristic of the learning game based on **U**. Consider the following stage-games:

$$\mathbf{u} = \begin{array}{c} \\ a \\ b \end{array} \begin{array}{c} a \quad\quad b \\ \begin{pmatrix} 8,9 & 2,3 \\ 0,1 & 4,5 \end{pmatrix} \end{array} \quad \tilde{\mathbf{u}} = \begin{array}{c} \\ a \\ b \end{array} \begin{array}{c} a \quad\quad b \\ \begin{pmatrix} 8,9 & 6,3 \\ 0,7 & 4,5 \end{pmatrix} \end{array} \quad \hat{\mathbf{u}} = \begin{array}{c} \\ a \\ b \end{array} \begin{array}{c} a \quad\quad b \\ \begin{pmatrix} 8,9 & 10,11 \\ 0,1 & 4,5 \end{pmatrix} \end{array}$$

We have $\mathbf{u}, \tilde{\mathbf{u}}, \hat{\mathbf{u}} \in \mathbf{CI}$.

Now consider

$$(u_R, \tilde{u}_C) = \begin{array}{c} \\ a \\ b \end{array} \begin{array}{c} a \quad\quad b \\ \begin{pmatrix} 8,9 & 2,3 \\ 0,7 & 4,5 \end{pmatrix} \end{array} \quad (u_R, \hat{u}_C) = \begin{array}{c} \\ a \\ b \end{array} \begin{array}{c} a \quad\quad b \\ \begin{pmatrix} 8,9 & 2,11 \\ 0,1 & 4,5 \end{pmatrix} \end{array}.$$

Both games have a unique and pure Nash equilibrium. Since **U** is a product class and $\mathbf{CI} \subseteq \mathbf{U}$, we have that both $(u_R, \tilde{u}_C), (u_R, \hat{u}_C) \in \mathbf{U}$.

Since $(a, a)$ is the unique Nash equilibrium of $(u_R, \tilde{u}_C)$ and $(b, b)$ is the unique Nash equilibrium of $(u_R, \hat{u}_C)$, the profile of learning heuristics $\boldsymbol{\sigma} = (\sigma_R, \sigma_C)$ must converge to $(a, a)$ in $(u_R, \tilde{u}_C)$ and $(b, b)$ in $(u_R, \hat{u}_C)$.

All of the above games are generic. Let $\tilde{\mathbf{U}}$ be an open neighborhood of $(u_R, \tilde{u}_C)$ and $\hat{\mathbf{U}}$ be an open neighborhood of $(u_R, \hat{u}_C)$ such that both are product classes and $\tilde{\mathbf{U}}, \hat{\mathbf{U}} \subseteq \mathbf{U}$ and retain the same pure best response structure, respectively.

Let $\tilde{\sigma}_C$ be an alternative learning heuristic for Colin that in games in $\tilde{\mathbf{U}} \cup \hat{\mathbf{U}}$ behaves identically to $\sigma$ in $\tilde{\mathbf{U}}$. For all other games, $\tilde{\sigma}$ behaves identically to $\sigma$. That is, with $\tilde{\sigma}_C$ Colin pretends to be in $\tilde{\mathbf{U}}$ whenever he is $\hat{\mathbf{U}}$. Note that $(\sigma_R, \tilde{\sigma}_C)$ converges to $(a, a)$ in stages-games in $\hat{\mathbf{U}}$, which is not the Nash equilibrium of these stage-games. Note further that Colin earns a strictly larger payoff in $(a, a)$ than in the unique Nash equilibrium of these games. Since any nonempty open neighborhoods must have strict positive Lebesgue measure, Colin's average long run payoff from $\tilde{\sigma}_C$ in the learning game when Rowena follows $\sigma_R$ is strictly larger than from $\sigma_C$. Thus, $\boldsymbol{\sigma}$ is not a Nash equilibrium of the learning game, a contradiction.

Note that we can extend the arguments above to games with larger actions sets by simply adding strictly dominated rows and columns.

Note further that we can let **U** be the smallest product class of games that contains **CI**, i.e.,

$$\mathbf{U} := \left\{ \mathbf{u} \in [0,1]^{2 \cdot |A^2|} : \begin{array}{l} \mathbf{u} = (\tilde{u}_R, \hat{u}_C) \text{ s.t. there exist } \tilde{\boldsymbol{u}} = (\tilde{u}_R, \tilde{u}_C) \\ \text{and } \hat{\boldsymbol{u}} = (\hat{u}_R, \hat{u}_C) \text{ with } \tilde{\boldsymbol{u}}, \hat{\boldsymbol{u}} \in \mathbf{CI} \end{array} \right\}.$$

Finally note that both games, $(u_R, \tilde{u}_C)$ and $(u_R, \hat{u}_C)$, posses a unique, pure, and strict Nash equilibrium. Thus, it must also be the correlated equilibrium, the iterative admissible action profile, the rationalizable action profile, and the minimal curb set. Hence, analogous arguments apply when we weaken convergence to the set of correlated equilibria, iterative admissible action profiles, rationalizable action profiles, or minimal curb sets, respectively. This completes the proof of Proposition 4. □

# References

[1] Aumann, R. (1974). Subjectivity and correlation in randomized strategies, Journal of Mathematical Economics 1, 67–96.

[2] Aumann, R. and S. Sorin (1989). Cooperation and bounded recall, Games and Economic Behavior 1, 5–39.

[3] Babichenko, Y. (2010). Uncoupled automata and pure Nash equilibria, International Journal of Game Theory 39, 483–502.

[4] Başar, T. and G.J. Olsder (1999). Dynamic non-cooperative game theory, 2nd edition, SIAM.

[5] Basu, K. and J. Weibull (1991). Strategy subsets closed under rational behavior, Economics Letters 36, 141–146.

[6] Bernheim, B.D. (1984). Rationalizable strategic behavior, Econometrica 52, 1007-1028.

[7] Biggs, J.B. (1985). The role of metalearning in study processes, British Journal of Educational Psychology 55, 185–212.

[8] Binmore, K. and L. Samuelson (1992). Evolutionary stability in repeated games played by finite automata, Journal of Economic Theory 57, 2780–305.

[9] Branfman, R.I. and M. Tennenholtz (2004). Efficient learning equilibrium, Artificial Intelligence 159, 27–47.

[10] Camerer, C. F., Ho, T.-H. and J.-K. Chong (2002). Sophisticated experience-weighted attraction learning and strategic teaching in repeated games, Journal of Economic Theory 104, 137–188.

[11] Caprara, G. V., Fida, R., Vecchione, M., Del Bove, G., Vecchio, G. M., Barbaranelli, C., and A. Bandura (2008). Longitudinal analysis of the role of perceived self-efficacy for self-regulated learning in academic continuance and achievement, Journal of Educational Psychology 100, 525–534.

[12] Chong, J.-K., Camerer, C. F., and T.-H. Ho (2006). A learning-based model of repeated games with incomplete information, Games and Economic Behavior 55, 340–371.

[13] Cripps, M., Schmidt, K., and J. Thomas (1996). Reputation in perturbed repeated games, Journal of Economic Theory 69, 387–410.

[14] Cripps, M. and J. Thomas (1995). Reputation and commitment in two-person repeated games without discounting, Econometrica 63, 1401–1419.

[15] Cripps, M. and J. Thomas (2003). Some asymptotic results in discounted repeated games of one-sided incomplete information, Mathematics of Operations Research 28, 433–462.

[16] Demichelis, S. (2013). Efficient coordination in repeated games: Behavioral maxims, mimeo.

[17] Duersch, P., Kolb, A., Oechssler, J. and B.C. Schipper (2010). Rage against the machines: How subjects learn to play against computers, Economic Theory 43, 407–430.

[18] Duersch, P., Oechssler, J., and B.C. Schipper (2012). Unbeatable imitation, Games and Economic Behavior 76, 88–96.

[19] Duersch, P., Oechssler, J., and B.C. Schipper (2014). When is Tit-for-tat unbeatable?, International Journal of Game Theory 43, 25–36.

[20] Ellison, G. (1997). Learning from personal experience: One rational guy and the justification of myopia, Games and Economic Behavior 19, 180–210.

[21] Foster, D. and R. Vohra (1997). Calibrated learning and correlated equilibrium, Games and Economic Behavior 21, 40–55.

[22] Foster, D. and H.P. Young (2001). On the impossibility of predicting the behavior of rational agents, Proceedings of the National Academy of Sciences 98, 12848–12853.

[23] Foster, D. and H.P. Young (2003). Learning, hypothesis testing, and Nash equilibrium, Games and Economic Behavior 45, 73–96.

[24] Foster, D. and H.P. Young (2006). Regret testing: learning to play Nash equilibrium without knowing you have an opponent, Theoretical Economics 1, 341-367.

[25] Fudenberg, D. and D. Levine (1999). Conditional universal consistency, Games and Economic Behavior 29, 104–130.

[26] Fudenberg, D. and D. Levine (1998). The theory of learning in games, MIT Press.

[27] Fudenberg, D. and D. Levine (1989). Reputation and equilibrium selection in games with a patient player, Econometrica 57, 759–778.

[28] Fudenberg, D. and E. Maskin (1990). Evolution and cooperation in noisy repeated games, American Economic Review Papers & Proceedings 80, 274–279.

[29] Germano, F. (2007). Stochastic evolution of rules for playing normal-form games, Theory and Decision 62, 311–333.

[30] Germano, F. and G. Lugosi (2007). Global Nash convergence of Foster and Young's regret testing, Games and Economic Behavior 60, 135–154.

[31] Hart, S. and A. Mas-Colell (2006). Stochastic uncoupled dynamics and Nash equilibrium, Games and Economic Behavior 57, 286–303.

[32] Hart, S. and A. Mas-Colell (2003). Uncoupled dynamics do not lead to Nash equilibrium, American Economic Review 93, 1830-1836.

[33] Hart, S. and A. Mas-Colell (2001). A General class of adaptive strategies, Journal of Economic Theory 98, 26–54.

[34] Hart, S. and A. Mas-Colell (2000). A simple adaptive procedure leading to correlated equilibrium, Econometrica 68, 1127–1150.

[35] Hyndman, K., Ozbay, E.Y., Schotter, A., and W.Z. Ehrblatt (2012). Convergence: An experimental study of teaching and learning in repeated games, Journal of the European Economic Association 10, 573–604.

[36] Israeli, E. (1999). Sowing doubt optimally in two-person repeated games, Games and Economic Behavior 28, 203–216.

[37] Jordan, J.S. (1991). Bayesian learning in normal form games, Games and Economic Behavior 3, 60–81.

[38] Kakade, S.M. and D.P. Foster (2008). Deterministic calibration and Nash equilibrium, Journal of Computer and System Sciences 74, 115–130.

[39] Kalai, E. and E. Lehrer (1993). Rational learning leads to Nash equilibrium, Econometrica 61, 1019–1045.

[40] Kim Y.K. (1994). Evolutionarily stable strategies in the repeated prisoner's dilemma, Mathematical Social Sciences 28, 167–197.

[41] Lipman, B. (1991). How to decide how to decide how to ...: Modeling limited rationality, Econometrica 59, 1105–1125.

[42] Mailath, G. and L. Samuelson (2006). Repeated games and reputations, Oxford University Press.

[43] Meggido, N. (1986). Remarks on bounded rationality, IBM Research, Almaden Research Center.

[44] Meggido, N. and A. Wigderson (1986). On play by means of computing machines, in: Halpern, J.Y. (ed.), Theoretical Aspects of Reasoning about Knowledge, Morgan Kaufmann Publishers, 259–274.

[45] Milgrom, P. and J. Roberts (1990). Rationalizability, learning, and equilibrium in games with strategic complementarities, Econometrica 58, 1255–1277.

[46] Monderer, D. and L.S. Shapley (1996). Potential games, Games and Economic Behavior 14, 124–143.

[47] Monderer, D. and M. Tennenholtz (2007). Learning equilibrium as a generalization of learning to optimize, Artificial Intelligence 171, 448–452.

[48] Mongin, P. and B. Walliser (1988). Infinite regressions in the optimizing theory of decision, in: Munier, B.E. (ed.), Risk, decision and rationality, D. Reidel Publishing Company, Dordrecht, 435–457.

[49] Moulin, H. (1979). Dominance solvable voting schemes, Econometrica 47, 1337–1351.

[50] Nachbar, J. (1997). Prediction, optimization, and learning in repeated games, Econometrica 65, 275–309.

[51] Nachbar, J. (2001). Bayesian learning in repeated games of incomplete information, Social Choice and Welfare 18, 303–326.

[52] Nachbar, J. (2005). Beliefs in repeated games, Econometrica 73, 459–480.

[53] Pearce, D. (1984). Rationalizable strategic behavior and the problem of perfection, Econometrica 52, 1029–1050.

[54] Rietveld C.A., Medland, S., Derringer, J., Authors from the Social Science Genetic Association Consortium, Visscher, P.M., Benjamin, D.J., Cesarini, D., and P.D. Koellinger (2013). GWAS of 126,559 individuals identifies genetic variants associated with educational attainment, Science 340, 1467–1471.

[55] Sadzik, T. (2011). Coordination in learning to play Nash equilibrium, mimeo.

[56] Schipper, B.C. (2009). Imitators and optimizers in Cournot oligopoly, Journal of Economic Dynamics and Control 33, 1981–1990.

[57] Schipper, B.C. (2011). Strategic control of myopic best reply in repeated games, University of California, Davis.

[58] Schmidhuber, J. (2006). Gödel machines: fully self-referential optimal universal problem solvers, in: B. Goertzel, B. and Pennachin, C. (Eds.), Artificial General Intelligence, Springer Verlag, 199–226.

[59] Shalev, J. (1994). Nonzero-sum two-person repeated games with incomplete information and known-own payoffs, Games and Economic Behavior 7, 246–259.

[60] Sorin, S. (1999). Merging, reputation, and repeated games with incomplete information, Games and Economic Behavior 29, 274–308.

[61] Terracol, A. and J. Vaksmann (2009). Dumbing down rational players: Learning and teaching in an experimental game, Journal of Economic Behavior and Organization 70, 54–71.

[62] Young, H.P. (2009). Learning by trial and error, Games and Economic Behavior 65, 626–643.

[63] Young, H.P. (2004). Strategic learning and its limits, Oxford University Press.

[64] Vega-Redondo, F. (1997). The evolution of Walrasian behavior, Econometrica 65, 375–384.